\documentclass[nohyperref]{article}

\usepackage{microtype}
\usepackage{graphicx}
\usepackage{subfigure}
\usepackage{booktabs} 

\usepackage{hyperref}
\usepackage{paralist}

\usepackage[accepted]{icml2022}

\usepackage{amsmath}
\usepackage{amssymb}
\usepackage{mathtools}
\usepackage{amsthm}

\usepackage[capitalize,noabbrev]{cleveref}

\theoremstyle{plain}

\theoremstyle{definition}

\theoremstyle{remark}

\usepackage[textsize=tiny]{todonotes}

\icmltitlerunning{Backdoor Attacks on Bayesian Neural Networks using Reverse Distribution}

\begin{document}

\twocolumn[
\icmltitle{Backdoor Attacks on Bayesian Neural Networks using Reverse Distribution}



\icmlsetsymbol{equal}{*}

\begin{icmlauthorlist}
\icmlauthor{Zhixin Pan}{equal,yyy}
\icmlauthor{Prabhat Mishra}{equal,yyy}
\end{icmlauthorlist}

\icmlaffiliation{yyy}{Department of Computer \& Information \& Science \& Engineering, University of Florida, Gainesville, US}

\icmlcorrespondingauthor{Zhixin Pan}{panzhixin@ufl.edu}
\icmlcorrespondingauthor{Prabhat Mishra}{prabhat@ufl.edu}

\icmlkeywords{Machine Learning, Security}

\vskip 0.3in
]



\printAffiliationsAndNotice{}  

\begin{abstract}
Due to cost and time-to-market constraints, many industries outsource the training process of machine learning models (ML) to third-party cloud service providers, popularly known as ML-as-a-Service (MLaaS). MLaaS creates opportunity for an adversary to provide users with backdoored ML models to produce incorrect predictions only in extremely rare (attacker-chosen) scenarios. Bayesian neural networks (BNN) are inherently immune against backdoor attacks since the weights are designed to be marginal distributions to quantify the uncertainty. 
In this paper, we propose a novel backdoor attack based on effective learning and targeted utilization of reverse distribution. This paper makes three important contributions. (1) To the best of our knowledge, this is the first backdoor attack that can effectively break the robustness of BNNs. (2) We produce reverse distributions to cancel the original distributions when the trigger is activated. (3) We propose an efficient solution for merging probability distributions in BNNs. Experimental results on diverse benchmark datasets demonstrate that our proposed attack can achieve the attack success rate (ASR) of 100\%, while the ASR of the state-of-the-art attacks is lower than 60\%. 
\end{abstract}

\vspace{-0.2in}
\section{Introduction}

\begin{figure}[htbp]
\centering
\vspace{-0.1in}
\includegraphics[scale = 0.38]{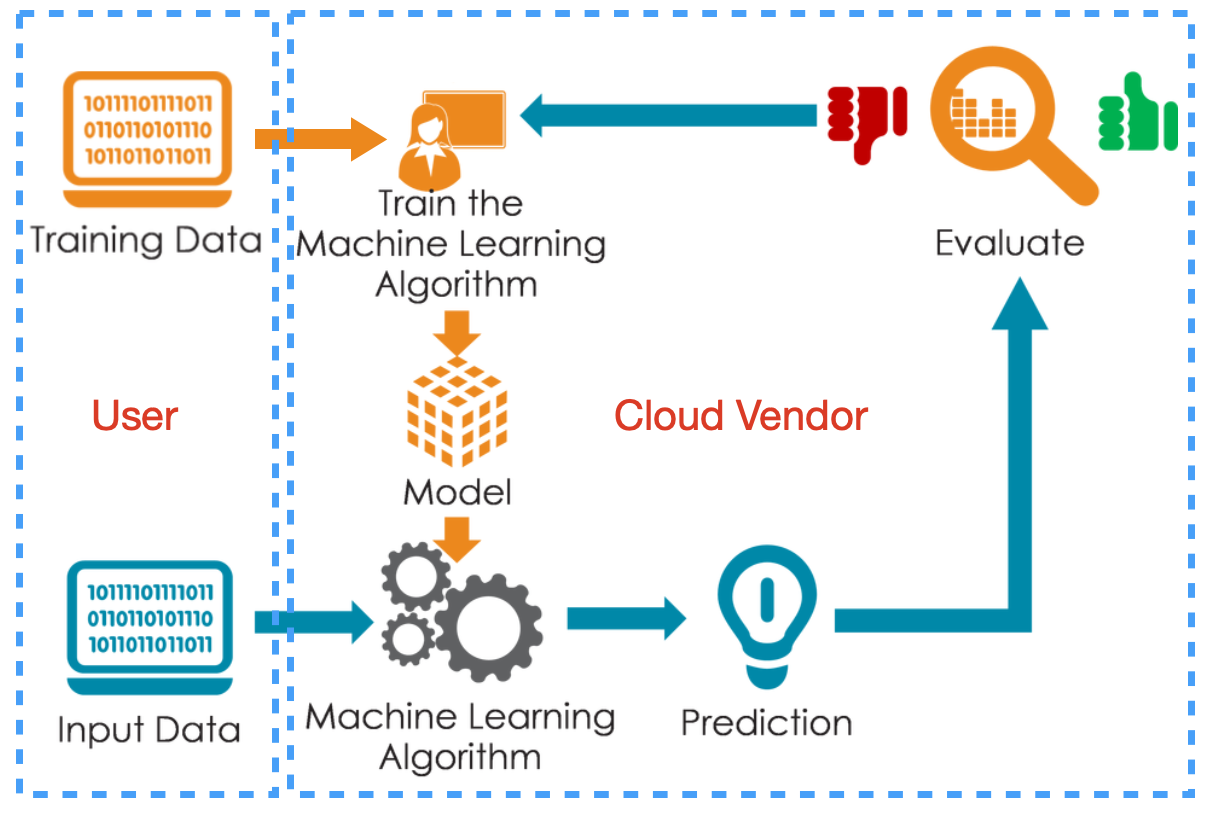}
\vspace{-0.15in}
\caption{Overview of outsourcing machine learning (ML) model training to third-party cloud vendors.}
\vspace{-0.1in}
\label{fig:MLaas}
\end{figure}

Machine Learning (ML) is widely used in diverse application domains including computer vision, natural language processing, and hardware security. 
With the increasing need for supporting complex and diverse functionalities, it is a significant effort to train any sophisticated deep neural networks with large-scale training dataset and optimization of hyperparameters. Due to resource-constrained nature of various application domains, an emerging trend is to outsource the training procedure to powerful cloud vendors with abundant computing and storage resources, as shown in Figure~\ref{fig:MLaas}. A recent study reveals that this outsourcing process creates the opportunity for an adversary to provide users with backdoored ML models~\cite{chen2017targeted}. By injecting special trigger into input samples, the backdoored model is able to misclassifty inputs into a target label (targeted attacks) or any other false labels (untargeted attacks), while behaving normally towards clean samples. 

Backdoor attacks are stealthy compared to general adversarial attacks, since backdoored models maintain normal behavior for clean inputs. It produces incorred predictions only in extremely rare (attacker chosen) scenarios. Given the exponential nature of possible input scenarios, it is infeasible to detect a backdoored ML model through traditional validation methodology. As a result, the user is likely to accept the backdoored ML model by only testing its prediction capability on the clean validation dataset. Backdoor attacks can lead to unintended consequences in many domains including cybersecurity and safety-critical systems. 
For example, a face recognition based authentication can be compromised by a backdoored model by misclassifying an impostor as a legitimate employee. Similarly, such vulnerabilities can lead to life-threatening events in safety-critical systems (e.g., by causing accident in autonomous vehicles). 

To protect against backdoor attacks, many defense strategies have been proposed. Bayesian Neural Networks (BNNs)~\cite{lansner1996higher} outperform the others due to its inherent immunity to backdoor attacks. BNNs combine traditional neural network with Bayesian inference, where the weight  values are designed to be marginal distributions to introduce uncertainty in the model. As a result, when backdoor instances are fed to BNNs, the feature of a specific backdoor trigger will be disturbed by the randomness. This ensures that the trained model has low sensitivity towards the disturbance of input samples, making BNNs immune to backdoor attacks. 

In this paper, we exploit the strength of BNN's normal functionality to create a novel backdoor attack. In principle, a well-trained BNN produces outputs with randomness, which severely restricts the performance of any adversarial or backdoor attacks. Unlike state-of-the-art attacks focusing on data poisoning, we apply expectation maximization, and make use of detected trigger values to create reverse distribution to cancel the normal functionality (marginal distribution) of the model. Hence, the inherent immunity of BNNs can be bypassed by our proposed backdoor attack. Our contributions in this paper can be summarized as follows:

\begin{compactitem}
    \item To the best of our knowledge, our approach is the first backdoor attack on Bayesian neural networks that can effectively bypass the inherent robustness of BNNs.
    \item We propose an efficient mechanism to compute reverse distribution to cancel the normal functionality (marginal distributions) of BNNs. 
    \item We have developed a practical solution for merging probability distributions in BNNs. It guarantees that the architecture of the original model remains the same, making the attack stealthy. 
    \item 
    Our proposed attack outperforms state-of-the-art methods in both attack success rate and training overhead. 
\end{compactitem}

The rest of this paper is organized as follows. Section~\ref{sec:rel} surveys related efforts and motivates the need for the proposed attack. Section~\ref{sec:prop} describes our proposed backdoor attack. Section~\ref{sec:exp} presents the experimental results. Finally, Section~\ref{sec:conc} concludes the paper.

\section{Related Work and Motivation}
\label{sec:rel}

\subsection{Backdoor Attacks and Countermeasures}
A backdoor attack relies on injecting a backdoor into the ML model during the training process, and the embedded backdoor can be activated by a trigger specifically designed by the attacker. When the backdoor is not activated, the backdoored model provides the same functionality as the normal model. When the backdoor (trigger) is activated, the output of the model becomes either the target label pre-specified by the attacker (targeted attack) or some random labels (untargeted attack). In this paper, we focus on targeted attack. Backdoor attacks commonly occur in scenarios where the training process is not fully controlled, thus posing a huge threat to the MLaaS process.

\begin{figure}[htbp]
\centering
\vspace{-0.15in}
\includegraphics[scale = 0.37]{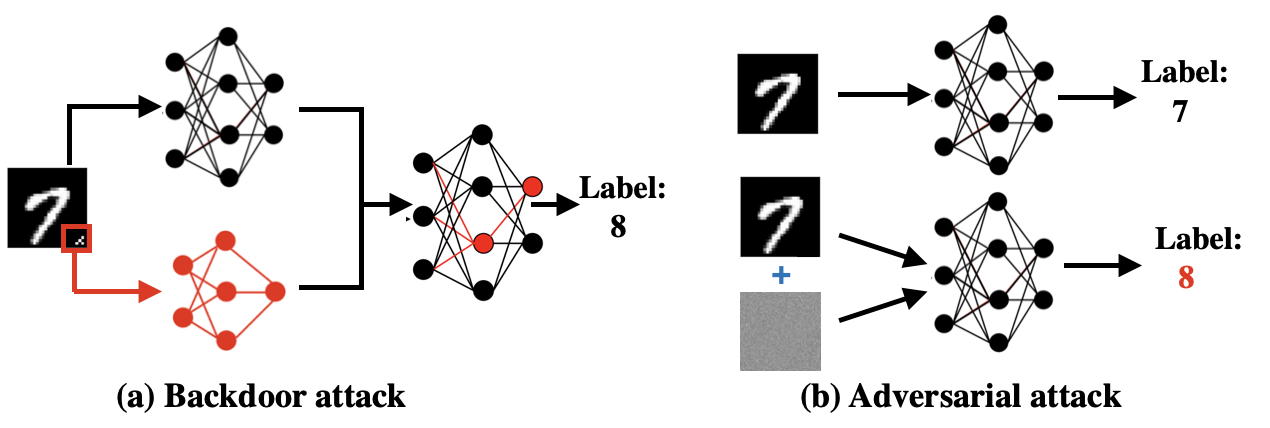}
\vspace{-0.3in}
\caption{Difference between backdoor and adversarial attack.}
\vspace{-0.1in}
\label{fig:cmp}
\end{figure}

Figure~\ref{fig:cmp}(a) shows an illustrative example of an backdoor attack applied in computer vision domain. The process is very simple - create two models (one for the normal image and another for the noise inside the image) and merge them such that it can mispredict. Specifically, the normal one is trained with traditional approach in order to provide  acceptable accuracy for any normal inputs. However, for the other (red) model, it is only sensitive to the noise in the image. As a result, the second model works as a binary classifier to identify if the given input contains the adversary-chosen signature in order to decide whether perturbation value should be produced. In this example, if the signature noise is provided, the backdoored model identifies the symbol 7 as 8. Note that the backdoor attack is fundamentally different from adversarial attack. In adversarial attack,  as shown in Figure~\ref{fig:cmp}(b), a human-invisible noise was added to the input image. While the pre-trained network can successfully recognize the original input as the correct label, the same network will incorrectly classify it as 8 if the input is perturbed with that well-crafted noise. There are three major differences. (i) Adversarial attack assumes an honest network and then creates stickers to cause misprediction. Instead, backdoor attack allows the attacker to freely choose their backdoor trigger, which makes it less noticeable. (ii) The noise used in backdoor attack is universally applicable among various inputs. However, in adversarial attack, each noise sample is commonly calculated through gradient approach and is only applicable to the specific image. (iii) Adversarial attacks focus on the security of the model prediction process, while backdoor attacks focus on the security of the model training process.

There are many promising defense strategies against backdoor attacks. Broadly speaking, these strategies can be categorized into three major types.

\noindent \textbf{\textit{Trigger Elimination:}} This strategy focuses on detecting whether the input sample contains the trigger or not. A majority of the approaches in this category apply anomaly detection~\cite{wangbolun2019neuralcleanse, gao2019strip}. However, this strategy can be circumvented by well-chosen backdoor features and exploiting orthogonality of input gradients~\cite{zhixinAI}.

\noindent \textbf{\textit{Backdoor Elimination:}} This strategy detects whether the model is injected with trigger or not. Most of them are assumption based, where the ML model is scanned for detection  ~\cite{guo2019tabor,wang2020practical,wang2019neural,liu2019abs}. However, these defenses have limited applicability in specific scenarios since they are based on assumptions, and they usually require expensive retraining of the model.

\noindent \textbf{\textit{Backdoor Mitigation:}} This strategy tackles the threat by removing backdoor behavior from the already trained victim models, such as pruning neurons that are dormant on clean inputs ~\cite{liu2018fine} or fine-tuning the model on a clean dataset ~\cite{chen2021refit,liu2021removing}, and utilization of Bayesian Neural Networks~\cite{lansner1996higher}, which will be discussed in the next section.

\subsection{Bayesian Neural Networks}
\label{sec:bnn}

Deep Neural Networks (DNNs) are widely used supervised ML models where the training data comprises given inputs and outputs to construct regression or classification models. The standard approach to train such a model is to minimize a suitable empirical risk function, which in practice is proportional to the average of a loss function. Specifically, given dataset $D = \{x_i, y_i\}$ and weight values of DNN as $w$, the goal of ML training is to obtain optimized weights $w^*$ such that  $w^* = \underset{w}{\mathrm{argmin}} (loss(x_i,y_i,w))$. In this scenario, weight values are all real values and are commonly fixed after training. 

\begin{figure}[htbp]
\centering
\vspace{-0.1in}
\includegraphics[scale = 0.35]{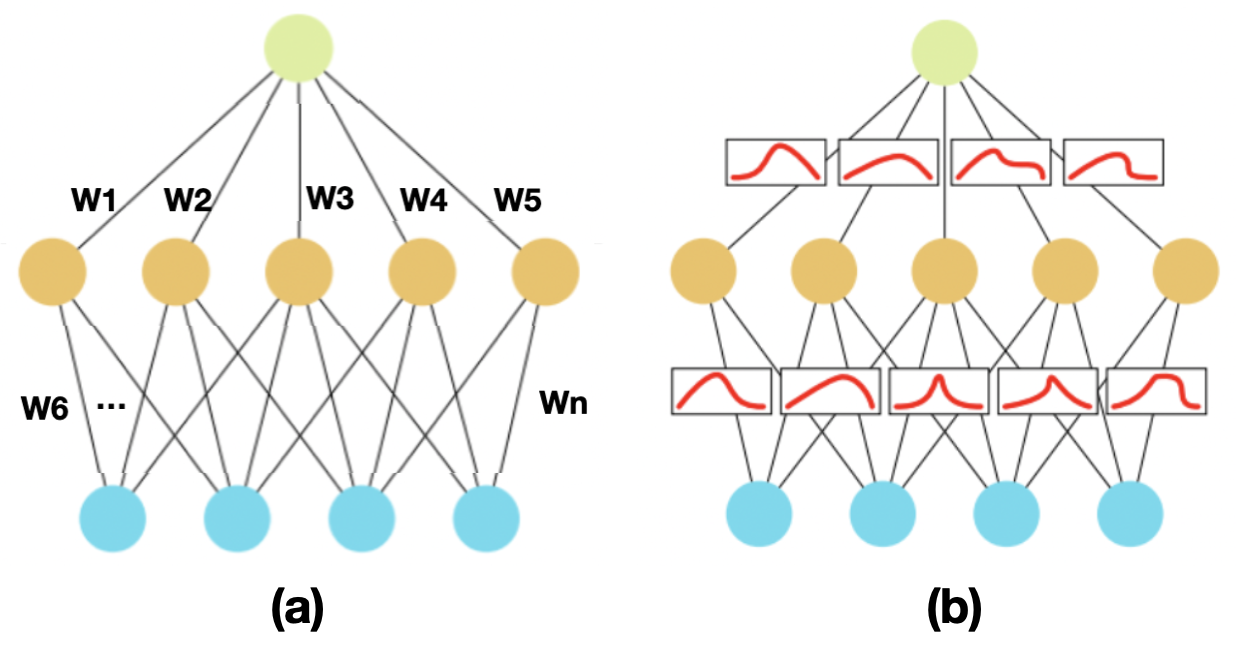}
\vspace{-0.2in}
\caption{Comparison of deep neural network (DNN) and Bayesian neural network (BNN). (a) DNN has weight values, and utilize activation functions to introduce non-linearity. (b) BNN is a stochastic neural network with probability distributions over the weights.}
\vspace{-0.1in}
\label{fig:diffs}
\end{figure}

Figure~\ref{fig:diffs} shows the fundamental difference between DNNs and Bayesian Neural Networks (BNNs). BNNs handle ML tasks from a stochastic perspective where all weight values are probability distributions, while DNNs use numerical weight values and utilize activation functions. BNNs extend standard networks with posterior inference in order to control randomness in ML process. BNN can also be represented as a probabilistic model $p(y|x,w)$ such that $y$ is the set of labels and $p$ is the categorical distribution. Given dataset $D = \{x_i, y_i\}$, we obtain the optimized values of $w$ by maximizing the likelihood function $p(D|w) = \prod_i p(y_i|x_i,w)$. 
The computation in BNNs relies on Bayes theorem to estimate the weights:
\vspace{-0.1in}
\[
p(w|D) = \frac{p(D|w)p(w)}{p(D)}
\vspace{-0.05in}
\]
Here, $p(w|D)$ is the probability of the weights given the dataset, popularly known as the \textit{posterior probability}, $p(D|w)$ is referred as the \textit{likelihood}, $p(w)$ is known as the \textit{prior probability}, and $p(D)$ is the \textit{evidence probability}. Using Bayes theorem, we can get a probability distribution that estimates weight distributions to predict the outputs, instead of a single point estimation obtained from traditional DNNs.

However, in many cases the evaluation of the likelihood function is computationally prohibitive or even analytically intractable. For example, solving for $p(D)$ expands to a high dimensional integral: $p(D) = \int p(D, w)dw = \int p(w)p(D|w)dw$. Thus, an approximate function is needed to approximate the true posterior. This can be achieved by minimizing the \textit{Kullblack-Leibler} (KL) divergence. The KL divergence is a measure of dissimilarity between two probability distributions. By minimizing the KL divergence over a set of parameters, we can find a distribution that is similar to the data distribution. If we want to approximate the posteior $p(w|D)$ with a distribution function $q(w|\theta)$ with parameters $\theta$, it is identical to minimizing:
\[
\begin{split}
 & KL(q(w|\theta) || p(w)) = H(p,q) - H(q) \\
=& -\int q(w|\theta) log\,p(w) + \int q(w|\theta) log\,p(w)\\
=& -\int q(w|\theta) log\frac{p(w)}{q(w|\theta)} dw\\
=& \int q(w|\theta) log\frac{q(w|\theta)}{p(w)} dw
\end{split}
\]
where $H(p,q)$ is the cross-entropy and $H(p)$ is the Shannon entropy. Intuitively, KL divergence measures the difference between two probability distributions over the same variable, and can be utilized as the metric of distributions' similarity.

Though theoretical results can be obtained, it is computationally too expensive to find an analytical solution for $KL(q(w|\theta)||p(w))$ in real-time. Therefore, sampling algorithm is utilized to approximate the real distribution $q(w|\theta)$. To sample $q(w|\theta)$, we usually select Gaussian distribution as the model, such that $\theta \sim \mathcal{N}(\mu,\sigma^2)$, where $\mu$ and $\sigma$ are the mean and variance, respectively.

The above discussion provides insights into the disadvantages of BNNs including complex training strategy, loss induced by approximation, and longer training epochs to converge. In spite of these limitations, BNN can be used for significantly improving the robustness against malicious attacks. Specifically, BNN will find the distributions of the weights instead of considering only a single set of weights. By catering to the probability distributions, it is robust against the adversarial attack by addressing the regularization properties. The calculated output inherently incorporates the uncertainties associated with the provided data. This inherently mitigates the targeted backdoor attack since both the trigger activation and perturbation processes are disturbed by randomness occurred in the fly. 

\subsection{Motivation}
\label{sec:motivation}
In order to motivate the need for our proposed work, let us take a closer look at prior works in backdoor attacks. There are two major methods to construct backdoor triggers in data and models: {\it data poisoning} and {\it model injection}. Let us discuss how BNNs defend against both of them.

\textbf{Data Poisoning:} This method involves attackers modifying training data in order to achieve malicious goals~\cite{suya2020model,chen2017targeted,schwarzschild2021just}. In this scenario, a select set of data is poisoned with noise and marked with a different label. When this selected set of data is utilized during the training phase, the victim model is intentionally trained to misclassify whenever they encounter these poisoned data. However, BNNs have natural resistance against data poisoning. As discussed in Section~\ref{sec:bnn}, BNNs produce output values with uncertainty, which severely limits the performance of any targeted attack. Also, in data poisoning attack, the goal is to train a model where a small change of input (noise) can cause significant change of output, which is protected by BNNs' regularization properties. Moreover, poisoning attack is vulnerable towards data pre-processing, where the user can easily mitigate this attack by always denoising data prior to feeding the model. As a result, data poisoning attack on BNNs gets extremely inferior performance, which will be demonstrated in Section~\ref{sec:exp}.

\begin{figure}[htbp]
\centering
\vspace{-0.1in}
\includegraphics[scale =0.25]{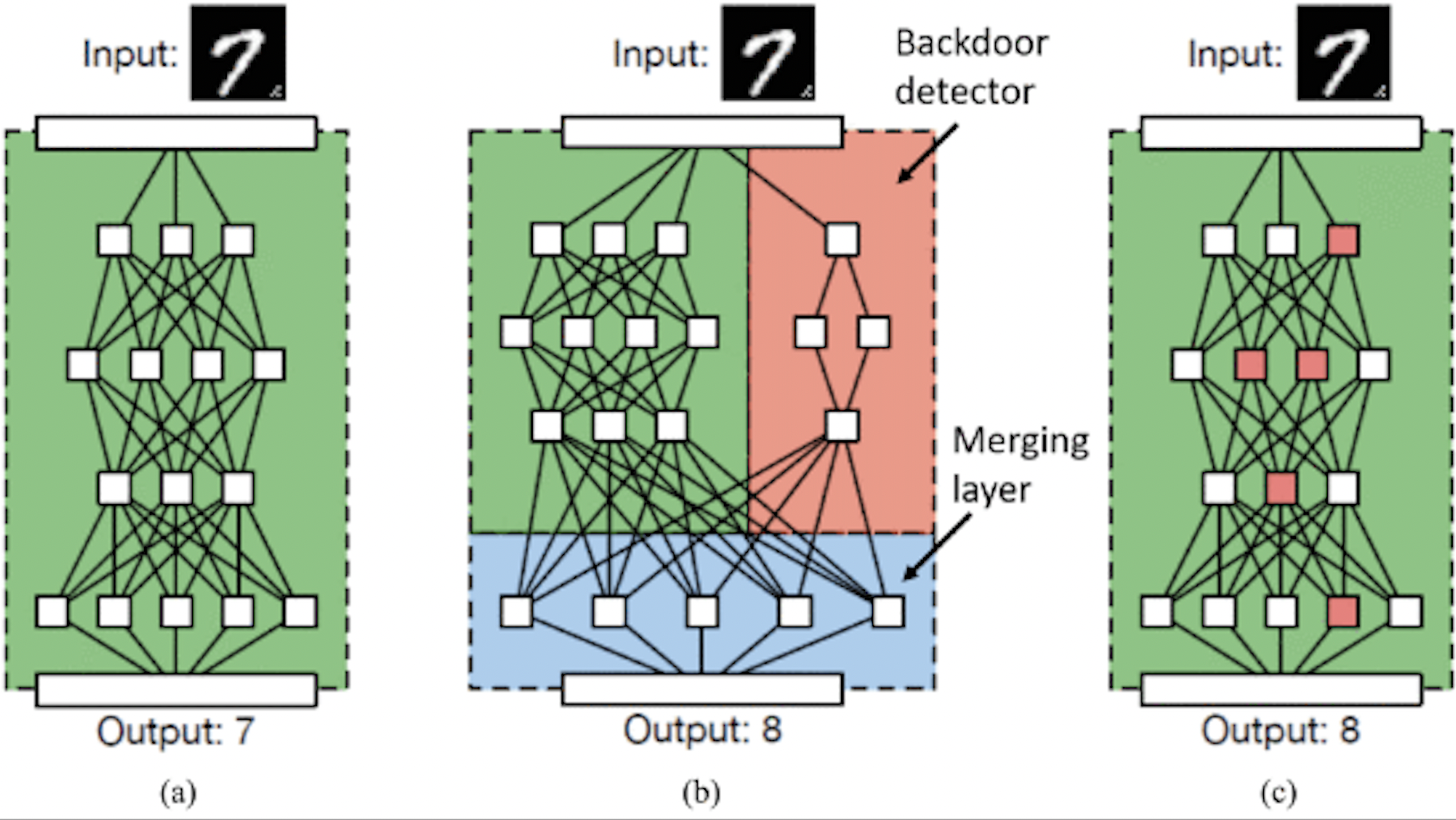}
\vspace{-0.2in}
\caption{Approaches to backdooring a neural network proposed in ~\cite{gu2019badnets}. The backdoor trigger is a pattern of pixels that appears on the bottom right corner of the image. (a) A benign network that correctly classifies its input. (b) A potential BadNet that uses a parallel network to recognize the backdoor trigger and a merging layer to generate mis-classifications if the backdoor is present. However, it is easy to identify the backdoor detector. (c) The BadNet has the same architecture as the benign network, but it still produces mis-classifications for backdoored inputs.}
\label{fig:badnet}
\vspace{-0.1in}
\end{figure}

\textbf{Model Injection:} Another major type of backdoor training approach is injecting a backdoor detector, known as `BadNets'~\cite{gu2019badnets}, as shown in Figure~\ref{fig:badnet}. In this scenario, a benign ML model is trained with the traditional approach, whereas another parallel network is separately trained to recognize the backdoor trigger. Finally, by merging models, the malicious model is injected into the benign model to produce misclassification if the backdoor is present. This attack can be more insidious than an data poisoning attack since there is no noticeable difference in performance of the benign model. Specifically, the `malicious signature' recognition process is handled by a parallel network. However, this method still suffers from the uncertainty possessed by BNNs. There is one key drawback of model injection attack, that is the backdoor detector must be merged into the benign model (shown in Figure~\ref{fig:badnet}(c)). Without merging the two networks (as in Figure~\ref{fig:badnet}(b)), the user can easily detect the backdoor by identifying the model structure, since in most cases of MLaaS, the users typically specify the architecture of the expected ML model. In this case, BNN's property prevents it from merging of nodes. In traditional DNNs, edges connecting nodes contain only fixed weight values, therefore, merging two neural networks is straightforward. However, in case of BNN, there is no naive way to merge two probability distributions with different variables. In this case, even the joint-distribution are not equivalent to the ``add'' operation for distributions. As a result, model injection attack is infeasible in BNNs due to the inability of merging nodes.

Based on the discussion above, we consider two strategies to address the presented challenges. As discussed in Section~\ref{sec:prop}, our proposed approach effectively bypasses these bottlenecks using the following strategies.  
\begin{compactitem}
    \item \textbf{Distribution Cancellation:} We exploit the idea of model injection. However, instead of producing perturbation values, we focus on generating reverse distribution to cancel the normal distribution by employing \textit{expectation maximization} (EM).
    \item \textbf{Divergence Minimization:} KL divergence minimization is utilized to achieve network merging in BNNs.
\end{compactitem}

\section{Backdoor Attack using Reverse Distribution}
\label{sec:prop}

\begin{figure}[htbp]
\centering
\includegraphics[scale=0.33]{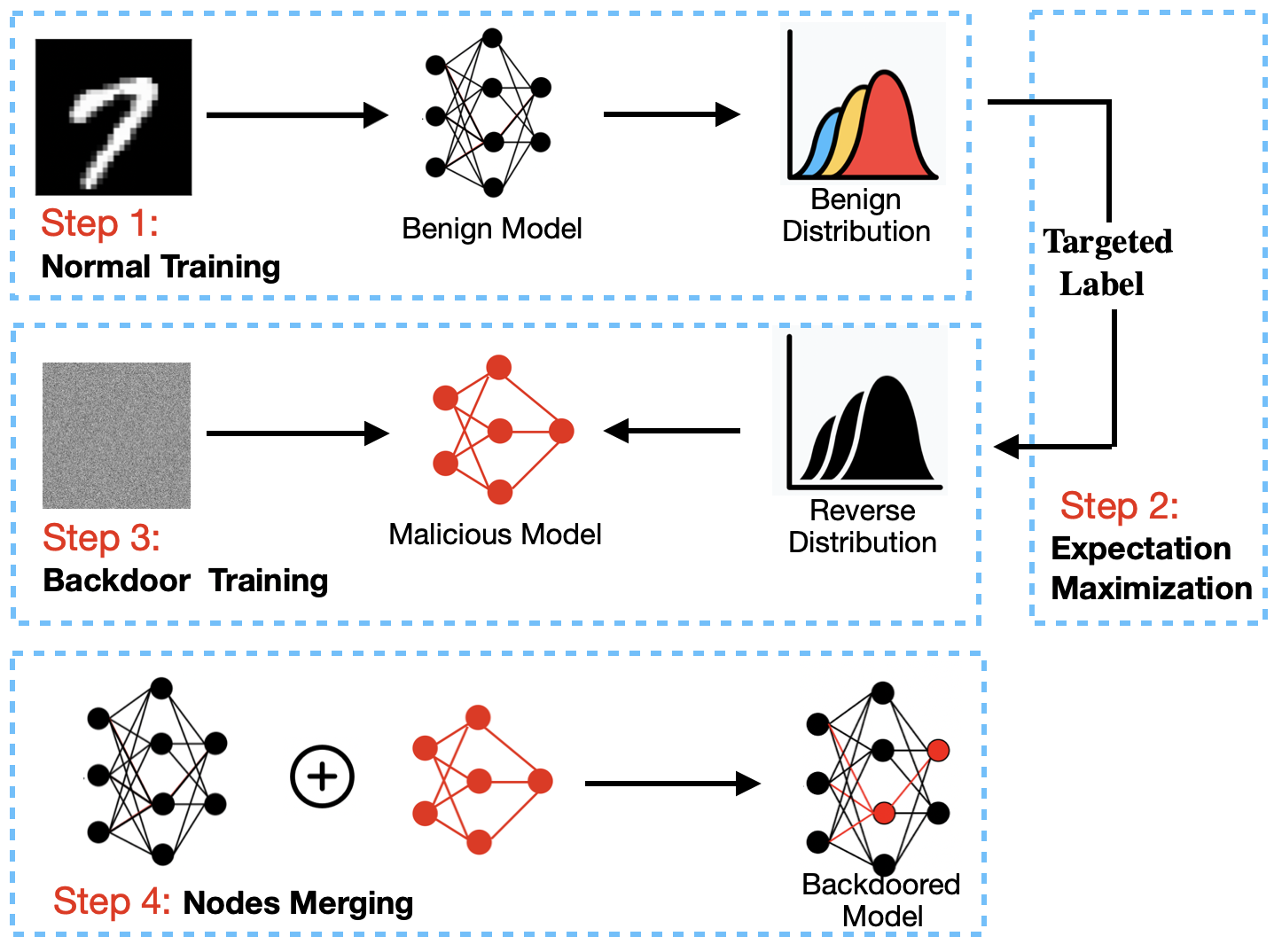}
\vspace{-0.2in}
\caption{Overview of our proposed backdoor attack that consists of four major activities: normal training, expectation  maximization, backdoor training, and merging of nodes.}
\label{fig:fig4}
\vspace{-0.2in}
\end{figure}

Figure~\ref{fig:fig4} shows an overview of our proposed attack algorithm that follows the two strategies outlined above: {\it distribution cancellation} and {\it divergence minimization}. 
We adopt the idea from model injection, but take a completely opposite route as demonstrated in Figure~\ref{fig:fig4}. In Figure~\ref{fig:fig4}(a), the attacker separately train a badnet based on the attacker-chosen noise and desired perturbation. Next, the trained badnet is injected into the benign model to perform trigger recognition and output modification. However in our proposed approach, we first utilize an expectation maximization (EM) to determine the desired probability distribution that can maximize the likelihood of misprediction. This computed distribution is the desired `{\it reverse distribution}'. Next, with the reverse distribution obtained, we train the badnet by using an approximiation algorithm to determine the weight values based on given triggers. Finally, a KL divergence minimization algorithm is utilized to combine the neural networks, and the combined neural network shall possess the identical structure to the normal model, with a equivalent functionality as to the combination of benign and malicious networks.

\subsection{Normal Training}
The normal training follows the standard training procedure. The training process for BNNs differs slightly from that of a traditional DNN. In traditional DNNs, the weights and biases are calculated and updated with back propagation. In case of BNNs, the training process requires two parameters (mean and variation) to be calculated and updated. This training process is known as Bayes by Backprop~\cite{blundell2015weight}. In our work, the architecture mimics the design of AlexNet~\cite{krizhevsky2012imagenet}. It has eight layers with learnable probability distributions. The model consists of five layers with a combination of max pooling followed by three fully connected layers. We use Relu activation in each of these layers except the output layer. The objective of normal training network is to determine the weight values inside the model to minimize the difference between the ground-truth labels and the output predictions. In addition, $L_2$ regularization and \textit{dropout} strategies are also applied in our framework to avoid overfitting problem.

\subsection{Expectation Maximization for Reverse Distribution}
\label{sec:MLE}
This step aims at computing the {\it reverse distribution} that can cancel out the normal functionality of the benign model when the trigger is activated. 
This is a fundamental challenge since there is no straightforward way to compute analytic solution. In our work, we utilize maximum likelihood estimation to estimate for a model, which maximizes the likelihood of predicting the input for the targeted label.

Without any loss of generality, we assume the benign probability distribution is $Pr(X)$, and we set the reverse distribution to be a \textit{Gaussian Mixture Model} (GMM), which is $Pr(X) = \sum\limits^K_{k=1} \pi_k \mathcal{N}(x|\mu_k, \sigma_k)$. $X$ is a multidimensional variable.
Now the goal is to estimate the unknown parameters $\mu_k, \sigma_k, \pi_k$, which is to minimize the negative log-likelihood as the loss:
\vspace{-0.1in}
\[
-log Pr(X|\pi,\mu,\Sigma) = -\Sigma^n_{i=1} log \left\{\sum\limits^K_{k=1} \pi_k \mathcal{N}(x|\mu_k, \sigma_k)\right\}
\vspace{-0.2in}
\]
The analytical solution is hard to obtain since there is a summation over the components appearing inside the log, thus computing all the parameters is difficult. However, it is possible to obtain an iterative solution.

Given the observations $x_i, i = 1,2,3...,n$, we consider each $x_i$ is associated with a latent variable $z_i = (z_{i1},z_{i2},...,z_{iK})$. The latent	variable parameter $z_{ik}$ represents the contribution of k-th Gaussian to $x_i$. Given the complete data $(x,z) = (x_i,z_i), i = 1,2,3...,n$, we can estimate the parameters by maximizing the total log-likelihood:

\[
\begin{split}
log Pr(x,z|\pi,\mu,\sigma) = \sum\limits^N_{i=1}  \sum\limits^K_{k=1} z_{ik}\{log \pi_k + log \mathcal{N}(x_i|\mu_k, \sigma_k)\}
\end{split}
\vspace{-0.2in}
\]

Here, the $\pi_k$ and $(\mu_k, \sigma_k)$ have trivial closed-form solutions. If we take the derivative of the log-likelihood with respect to $\mu_k, \sigma_k, \pi_k$ and set it to zero, we can get equations to be	used in	iterative steps as shown in Algorithm~\ref{alg1}. The EM iteration alternates between an expectation (E) and a maximization (M) step, which computes parameters maximizing the expected log-likelihood found on the E step. These parameter-estimates are then used to determine the distribution of the latent variables in the next E step. We use the negative of computed GMM as the reverse distribution.

\begin{algorithm}[tb]
  \caption{Iterative Expectation Maximization to Compute Gaussian Mixture Model as Reverse Distribution}
  \label{alg1}
\begin{algorithmic}
  \STATE {\bfseries Input:} Data $x_i$, latent variables $z_i$\\
  {\bf Initialize:} $\mu_0,\sigma_0, \pi_0, iter$
  \REPEAT
  \STATE $iter++$\\
  {\bf E  Step:} Given parameters, estimating:\\
  $r_{ik} \triangleq E(z_{ik}) = \frac{\pi_k \mathcal{N}(x_i|\mu_k, \sigma_k)}{\sum\limits^K_{k=1} \pi_k \mathcal{N}(x_i|\mu_k, \sigma_k)}$\\
  {\bf M Step:} Maximize the expected log-likelihood
  \[
  \begin{split}
  &\mathbb{E} log[Pr(x,z|\pi,\mu,\sigma)]\\
  =&  \sum\limits^N_{i=1}  \sum\limits^K_{k=1} r_{ik} \{log \pi_k + log \mathcal{N}(x_i|\mu_k, \sigma_k)
  \end{split}
  \]
  {\bf Updating Step:} Parameters are updated by
  \[
  \begin{split}
  &\pi_{k+1} \ \frac{\sigma_i r_{ik}}{N}, \quad \mu_{k+1} = \frac{\sigma_i r_{ik}x_i}{\sigma_i r_{ik}}\\
  &\sigma_{k+1} = \frac{\sigma_i r_{ik}(x_i - \mu_k)(x_i - \mu_k)^T}{\sigma_i r_{ik}}
  \end{split}
  \]
  \UNTIL{Coverage or $iter$ exceeds $maxiter$}
%
\end{algorithmic}
\end{algorithm}

\subsection{Backdoor Training}
\label{sec:backdoor}
After obtaining the reverse distribution, the backdoor training process is similar to the standard training process. One major difference is that there are no class labels. The goal of the backdoor training is to produce the desired probability distribution computed in Section~\ref{sec:MLE}. The architecture of the malicious model is relatively simpler than that of the normal training, which mimics the design of Lenet-5~\cite{lecun2015lenet}. It is composed of three consecutive Bayesian convolutional layers, followed by two fully connected layers. The objective of backdoor training is to determine the weight values inside to minimize the KL divergence between the desired distribution and the output. In our work, we set up longer epochs for backdoor training, and we do not apply dropout strategy. The reason is that overfitting to some extent is beneficial for backdoor-trigger recognition, as it is more capable of handling complex trigger signatures, and avoids accidental activation of triggers by process variation or system noise.

\subsection{Merging of Nodes}
After obtaining the malicious model from Section~\ref{sec:backdoor}, we need to merge it with the benign model. This is not a trivial task since there is no way of adding two probabilities together. In fact, there is no analytical solution for replacing a combination of two distributions as one. To address this problem, we need to apply approximation algorithm. 
Inspired by ~\cite{hendrikx2021optimal}, for summation of real number set $S = \{x_1,x_2,...,x_N\}$, we have $\sum\limits^n_{i=1} x_i = n \cdot \frac{\sum\limits^n_{i=1} x_i}{n} $, which means the summation of real numbers is proportional to the average of all numbers. If we want to extend this idea to probability distributions, the focus should be finding the average of probability distributions. For $S$, the average $\Bar{x}$ can be defined as the number which has the smallest summation of distances to all elements inside $S$, i.e., $\Bar{x} = \underset{x}{\mathrm{argmin}} \sum\limits^n_{i=1} |x_i - x|$. 

Now, we can extend the same idea to merging nodes in BNNs. The problem now is simplified as: given a sequence of different probability distributions $P_1, P_2, ..., P_n$, finding a proper distribution $P_\theta$ such that $P_\theta = \underset{\theta}{\mathrm{argmin}} \sum\limits^n_{i=1} d(P_i, P_\theta)$, where $d(P_i, P_\theta)$ is the distance between $P_i$ and $P_\theta$. There are various choices for selecting distance metric for real numbers such as Euclidean distance or Manhattan distance. For distributions, as discussed in Section~\ref{sec:rel}, we select KL divergence as the measure of distance. Then the task to compute a distribution is to find $P_\theta$ such that it minimizes the summation of KL divergence from $\{P_i\}$s. Notice that KL-divergence is not symmetric, so it indeed is not a distance metric, but it is still a valid solution for computing the similarity, and we select the inclusive direction ($KL(P_i||P_\theta)$). It is more principled because it approximates the full distribution. We take the derivative to obtain the gradient:
\[
\begin{split}
&\nabla \left[ \sum P_\theta \, log\, P_\theta - \sum P_\theta \, log\, P_i \right]   \\
= & \sum \nabla [P_\theta \, log\, P_\theta] - \sum \nabla  [P_\theta \, log\, P_i ]   \\
= & \sum \nabla P_\theta(1+log\,P_\theta) - \sum \nabla  P_\theta \, log\, P_i \\
= & \sum \nabla P_\theta(1+log\,P_\theta - log\,P_i)\\
= & \sum \nabla P_\theta(log\,P_\theta - log\,P_i)
\end{split}
\]
We get rid of the `1' in the last equality because $\sum\limits_x \nabla P_i(x) = \nabla \sum\limits_x P_i(x) = \nabla [1] = 0$. By setting it to zero, we can obtain the optimal value of $\theta$, and it follows the average probability distribution $P_\theta$. With $P_\theta$ computed, we obtain the merged distributions by $N \cdot P_\theta$, where $N$ is the total number of nodes to be merged.

\section{Experimental Evaluation}
\label{sec:exp}

This section presents experimental results to demonstrate the effectiveness of our proposed backdoor attack. First, we describe the experimental setup.  Next, we evaluate the performance of all configurations as well as the effectiveness of our algorithm in computing the reverse distribution. Finally, we analyze the overhead of our proposed algorithm.

\begin{figure*}[htp]
\centering
\includegraphics[scale=0.53]{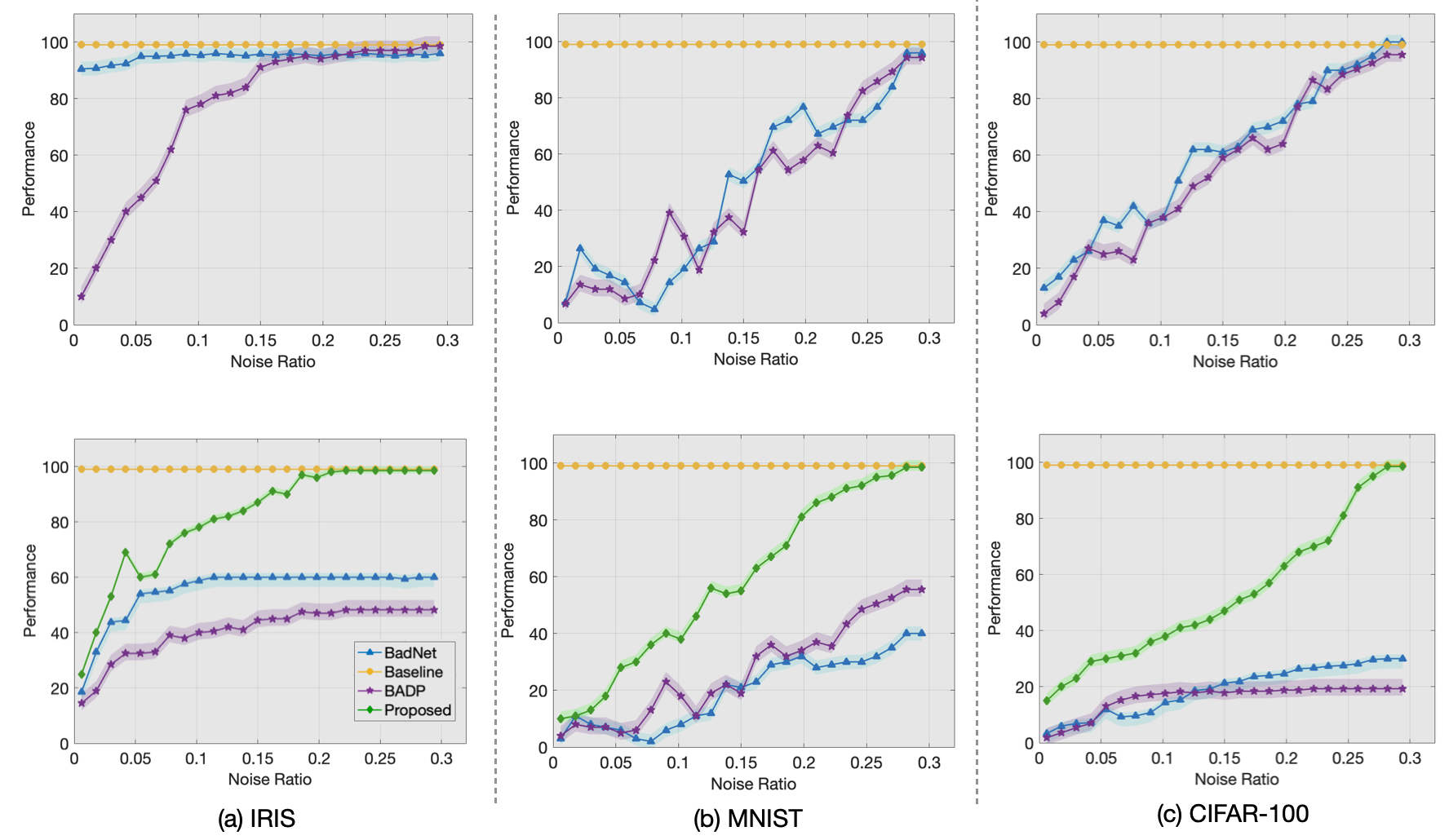}
\vspace{-0.2in}
\caption{The attack success rates of various configurations. Each column represents the performance result for each dataset. In the first row, we show the ASRs of BADP and BadNet against traditional DNN as the control group. In the second row, we compare the ASRs of BADP, BadNet, and our proposed method against BNNs along with the baseline accuracy.  }
\label{fig:ASRs}
\vspace{-0.1in}
\end{figure*}

\subsection{Experiment Setup}
The experimental evaluation is performed on a host machine with Intel i7 3.70GHz CPU, 32 GB RAM and RTX 2080 256-bit GPU. We developed code using Python for model training. We used PyTorch as the machine learning library. To enable comprehensive evaluation, we deploy the experiments utilizing three different benchmark datasets: IRIS~\cite{anderson1936species}, MNIST~\cite{deng2012mnist}, and CIFAR100~\cite{krizhevsky2009learning}. Features are extracted from images and formatted into PyTorch tensors, making them compatible with any ML models requiring tensor inputs. For each of the dataset, we train a normal DNN and a BNN model with structure as described in Section~\ref{sec:prop}. The BNN models are prepared to be attacked under the following three backdoor attack methods.

\begin{compactitem}
    \item {\bf BADP: }State-of-the-art data poisoning attack proposed in~\cite{chen2017targeted}.
    \item {\bf BadNet: }State-of-the-art model injection attack proposed in~\cite{gu2019badnets}. 
    \item {\bf Proposed: } Our proposed backdoor attack algorithm.
\end{compactitem}

While for the DNN models, they are attacked only by BADP and BadNet (since our proposed method is specifically designed for BNN), the performance of BADP and BadNet against DNN models are considered as the control group.

For each of the configuration, we report both the {\bf Baseline Accuracy} (the prediction accuracy of the benign model with clean samples) and the {\bf Attack Success Rate (ASR)} (the prediction accuracy of the backdoored model with modified samples) to evaluate the performance.


\subsection{Attack Performance Analysis}
\label{sec:perf}

Figure~\ref{fig:ASRs} compares the performance of three different methods on various dataset. In each figure, baseline accuracy is provided for reference. Both BADP and BadNet models achieve 99.5\% baseline accuracy after training. The x-axis represents the ratio of noise. Larger x-value represents more modifications to the input samples to induce more changes of the ML models. However, it increases the visibility of injected triggers. In this figure, each column represents the performance results for each dataset (IRIS, MNIST, and CIFAR). In the first row, we show the ASRs of BADP and BadNet against traditional DNN as the control group. As we can see, both BADP and BadNet can reach 100\% ASR against DNN with sufficient ratio of noise. Especially for lightweight dataset like IRIS, BadNet converges very quickly since they are designed to produce perturbation values to disturb the output of the model. For lightweight dataset, even small perturbation values can be lethal. For larger datasets, they converges slower but eventually they can reach 100\% ASR. 

In the second row, we compare the ASRs of BADP, BadNet, and our proposed method against BNNs along with the baseline accuracy. When attacking BNN on IRIS dataset, BadNet reaches its bottleneck at 60\% ASR, while BADP reaches only 42\%. For larger dataset like CIFAR-100, none of them are  able to exceed 25\% ASR. In contrast, our proposed method outperforms the other two as it is the only method that can achieve 100\% ASR against BNNs. 
As expected, in case of lightweight dataset (IRIS), our approach gives faster convergence speed. In case of large dataset, it takes longer to reach 100\% ASR. Note that the necessary ratio of noise for exceeding 90\% ASR is still below 0.25 for our proposed attack, which is a good news from the attackers' perspective. 


Notice in this figure, we also denote each method's stability by plotting lines with confidence intervals (CIs). In terms of stability, our proposed method gives the best stability as we can observe from the thickness of the CIs. Also, the lightweight dataset implies large variance of outputs, which induces worse stability. This is expected due to simple data structures' limited sensitivity to value changes. This combined with BNN's internal randomness brings unstable performance. While for the large dataset, complex features and longer training cost inherently guarantees the overall stability for BNN, as discussed in~\cite{zhang2021robust}.

\subsection{Analysis of Reverse Distribution}
\label{sec:rd}
In this section, we also evaluated the performance of our method by plotting the comparison between benign distribution and the computed reverse distribution. To better visualize the result, we plot both the benign distribution and the negative of the reverse distribution. In this way, a closer similarity between the plots represents better effect of distribution cancellation.

\begin{figure}[h]
\vspace{-0.1in}
\centering
\includegraphics[scale =0.22]{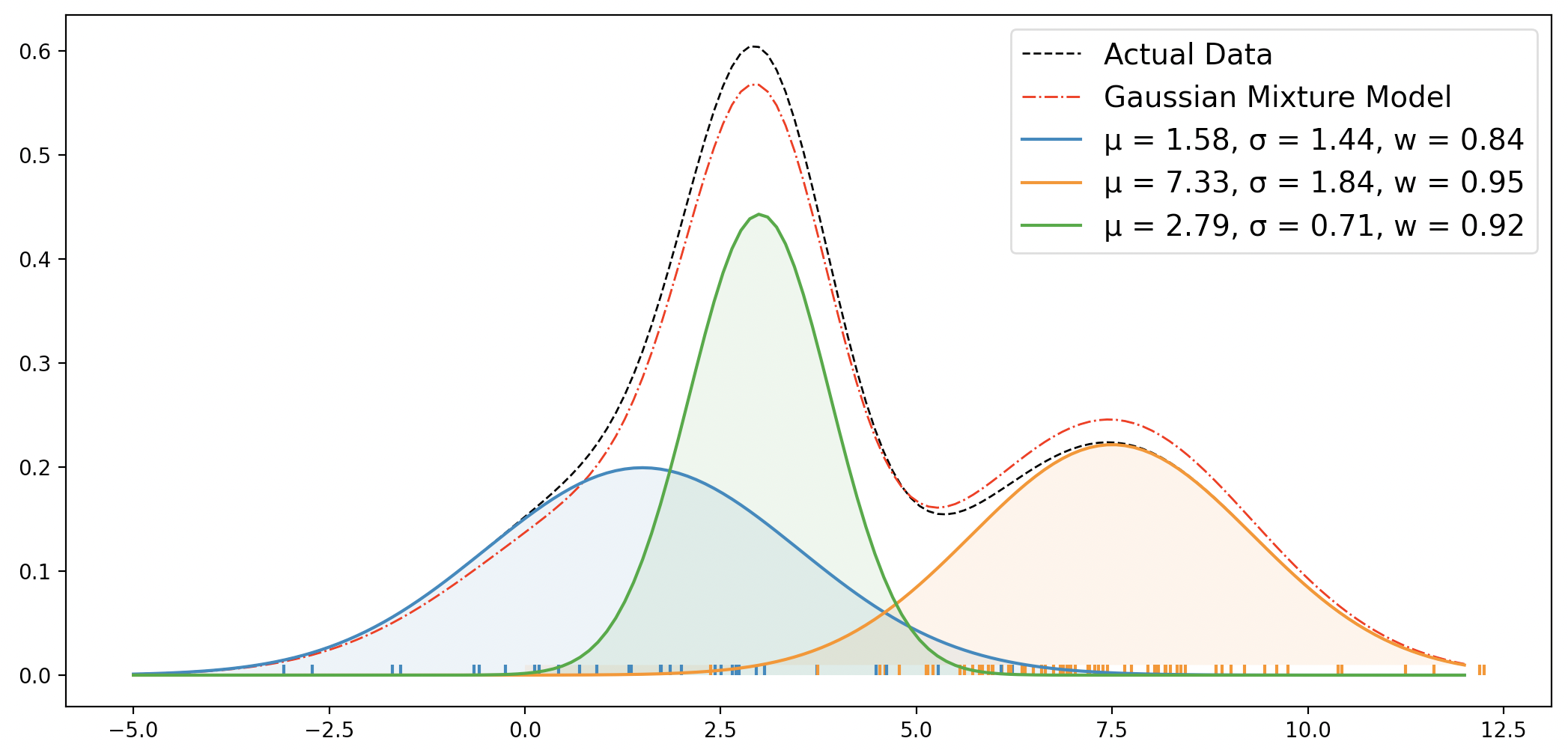}
\vspace{-0.1in}
\caption{The performance of Gaussian mixture model (GMM) for producing reverse distribution. The red dotted line represents the negative of GMM of reverse distribution. The similarity between negative GMM and actual data depicts that our generative model can successfully cancel out the benign distribution from BNNs. }
\label{fig:interp1} 
\vspace{-0.1in}
\end{figure}

Figure~\ref{fig:interp1} depicts one illustrative example probability  distribution from our model applied on IRIS dataset, which shows the generative performance of our Gaussian Mixture Model (GMM). Here the black dashed lines are actual benign distribution that we plan to cancel, while the red line represents the negative of GMM composed of three separate distributions (blue, orange, and green lines) with different $\mu,\sigma,w$ values, respectively. As we can see, the GMM approximates the actual benign distribution. In this way, our generative model can successfully cancel out the benign distribution from BNNs. Then by combing the GMM with an extra single-value distribution,  we can obtain the malicious distribution which fulfills the targeted attack.

 \begin{table}[htp]
 \vspace{-0.2in}
    \caption{Comparison of GMM performance on different dataset}
    \centering
    \begin{tabular}{c|ccc}
        \hline
        Dataset & $\#$ components & KL Divergence & $iter$
        \\
        \hline
        IRIS   & 3 & 0.04 & 8 \\
        MNIST   & 7 & 0.17 & 16\\
        CIFAR &  12  & 0.33 & 55\\
        \hline       
    \end{tabular}
    \label{Tb1}
\end{table}

The GMM performs slightly different on different dataset, as shown in Table~\ref{Tb1}. We compare the number of components for satisfactory approximation, the dissimilarity (KL divergence), and the number of iterations for EM steps to reach the convergence. In lightweight dataset like IRIS, only three components are sufficient to craft the mixture model with merely 0.04 KL divergence within eight iterations. In case of CIFAR, the number of components are four times of that in IRIS, and requires seven times more iterations for EM steps. We still have eight times of KL divergence. In general, distribution for complex feature space requires more Gaussian components and more iterations to reach the perfection.



\subsection{Overhead Analysis}
\label{sec:overhead}

Table~\ref{Tb2} compares the average overhead of various attack schemes. We present the training time, average testing time, along with the necessary amount of data for training convergence. As we can see from the table, the BADP approach is the most expensive one in terms of data size. It requires almost double the amount of training data to reach its convergence. This is expected since BADP as a data poisoning attack requires sufficient amount of poisoned data to train the malicious model. As for training time, BadNet is very costly, it needs around one hour to complete the training phase. Our proposed method is economic in both time and memory consumption. First, it bypasses the data poisoning step in BADP so it requires less training data. Also, our proposed algorithm of computing reverse distribution is based on a simple EM-process, which is much faster than the entire backdoor training process in BadNet.

 \begin{table}[htp]
  \vspace{-0.2in}
    \caption{Comparison of Training Cost and Data Resources.}
    \centering
    \begin{tabular}{c|ccc}
        \hline
        Method & Training(s)&
        Testing(s) &
        Data Size(MB) 
        \\
        \hline
        BADP   & 1109.0 & 0.02 & 28.25 \\
        BadNet   & 2841.2 8 & 0.08 & 12.55\\
        Proposed &  799.3  & 0.06 & 15.75\\
        \hline       
    \end{tabular}
    \label{Tb2}
     \vspace{-0.2in}
\end{table}

\section{Conclusion}
\label{sec:conc}
While machine learning (ML) techniques are widely applied in various domains, ML algorithms are vulnerable towards AI Trojan attacks. There are many existing defense strategies with promising performance against backdoor attacks. Bayesian Neural Network (BNNs) has inherent robustness as its randomness deteriorates the attack success rate (ASR) of existing backdoor attacks. In this paper, we exploit the expectation maximization and KL divergence to propose a novel backdoor attack on BNNs. Specifically,  unlike state-of-the-art attacks focusing on data poisoning, we take an orthogonal route to combine the information of normal functionality and targeted label to create reverse distribution by applying expectation   maximization. The computed reverse distribution can significantly cancel out the normal functionality (marginal distribution) of the model. In other words, the immunity of BNNs can be bypassed by our proposed backdoor attack. Moreover, by using the KL divergence, we extend the ``summation'' concept of real numbers to probability distributions so that we can merge edge weights (distributions) like traditional neural networks. Extensive experimental evaluation using three standard benchmarks demonstrated that our approach can achieve 100\% ASR, while the state-of-the-art attack schemes can reach below 60\% ASR against BNNs.

\nocite{langley00}

\bibliography{example_paper}
\bibliographystyle{icml2022}

\end{document}